# The Rising Dominance of Methods Across Science

Alexander Krauss
Spanish National Research Council, Barcelona, Spain
London School of Economics, London, UK
alexander.krauss@iae.csic.es, a.krauss@lse.ac.uk

Ariel Rosenfeld
Bar-Ilan University, Tel Aviv, Israel
ariel.rosenfeld@biu.ac.il

Lutz Bornmann
Science Policy and Strategy Department
Max Planck Society, Munich, Germany
bornmann@gv.mpg.de

*Abstract.* Scientific progress is traditionally narrated through the interplay of theoretical insights and experimental findings. Yet this view of science underplays a third and central pillar of progress: the methods that underlie both conceptual advances and empirical evidence. By analysing more than 3 million articles across science published between 1980 and 2019, we find that science has undergone a fundamental structural transition. The share of papers that primarily contribute new methods—*methods papers*—has doubled across science over the past four decades, rising universally across disciplines and citation impact levels. Rather than a gradual evolution, this transition marks a pivotal shift beginning in the early 1990s, aligning with the computational revolution and the emergence of data-intensive science. The surge in methodological research is not confined to the most cited, elite publications; it spans the full spectrum of scientific output. These findings reveal a systemic reorientation of the scientific ecosystem where reusable methods increasingly serve as the essential infrastructure of scientific advances, challenging the traditional dichotomy of theory and experimental research. As science becomes increasingly methods-driven, our results call for rethinking how research is evaluated, funded and organised—towards better incentivising method innovations. This is especially the case as expanding AI must be effectively integrated with scientific instruments to realise its full potential.

# 1 Introduction

Science is often narrated as a sequence of theoretical paradigms and empirical breakthroughs. But behind theoretical and experimental advances lies an often overlooked foundation: the methods driving those advances. Whether sequencing DNA, amplifying genes, modelling complex systems or measuring molecular structures, methods shape what scientists can measure and experiment. In fact, the most influential papers in modern science, as measured by citation impact, are increasingly dominated not by theoretical or experimental advances, but by advances in methods that underlie them (1, 2). The polymerase chain reaction (PCR) method (3), DNA sequencing techniques (4) and the Lowry protein analysis method (5) for example transformed biology—by enabling countless studies that apply them. These classic 'super-cited' *methods papers*—publications that develop a novel technique, algorithm, procedure, tool or instrument—often overshadow the very advances they enabled (6).

But do these classic, high-impact methods papers represent isolated anomalies, or do they reflect a broader structural reconfiguration across science? In other words: is science becoming more methods-centric? To address this question, we move beyond the super-cited outliers to analyse the structural role of methods papers across the entire range of scientific impact. Using a cross-validated machine-learning classifier applied to over 3 million papers in Web of Science (WoS, Clarivate) and Scopus (Elsevier) between 1980 and 2019, we identify and track the prevalence of methods-focused research across time, disciplines and citation strata (see Methods and data section). This enables analysing how the landscape of scientific knowledge is shifting. We first examine the prevalence of methods papers and whether they extend beyond the scientific elite to 'ordinary science' that constitutes the bulk of scientific literature. We then trace the evolution of methods papers over time and across different citation strata to determine whether and when a structural shift occurred. We finally assess the universality of the transition across diverse disciplines.

Our analysis reveals that science has undergone a major structural transition. We find that the share of methods papers doubled over the past four decades, rising from roughly 20% to 40% of all published research. This shift is visible not only among the most cited papers—where methods papers have long been dominant—but across the full distribution of scientific impact, including the vast majority of papers that receive less attention. Viewing this 'methodization' of science solely through the lens of singular methodological triumphs obscures this more fundamental transformation. This rise of methods is not a gradual evolution but marks a pivotal transition in the early 1990s, coinciding with the computational revolution and the emergence of data-intensive science. While the precise timing varies, the dominance and reorientation towards methods-focused research is universal across disciplines, from natural sciences to engineering and technology.

Important implications emerge. Conceptually, these findings challenge the traditional dichotomy between theory and empirical research: they underscore that widely reusable methods—which can often transfer across fields—have become essential infrastructure to achieve scientific advances. Strategically, the scientific community needs to recalibrate its attention and resources to value the methodological infrastructure underlying scientific advances. Science policy, funding mechanisms and career structures need to increasingly incentivise methodological innovation—particularly in light of expanding AI that needs to be effectively integrated with scientific tools to realise its full potential.

# 2 Results

**Beyond the scientific elite.** To determine whether the prominence of methods is solely a characteristic of 'super-cited' outliers, we stratified over 3 million papers by citation impact. We find that methods papers are not confined to the scientific elite, but are ubiquitous across the entire scientific landscape (Fig. 1). Across both WoS and Scopus datasets, methods papers consistently represent between 30% and 50% of publications at every citation level—whether they belong to the top 1% of cited papers or to the broad base of everyday research. This challenges the view that method papers are only significant when they become ‘citation classics’ but they actually represent a foundational component of research output at every level of impact.

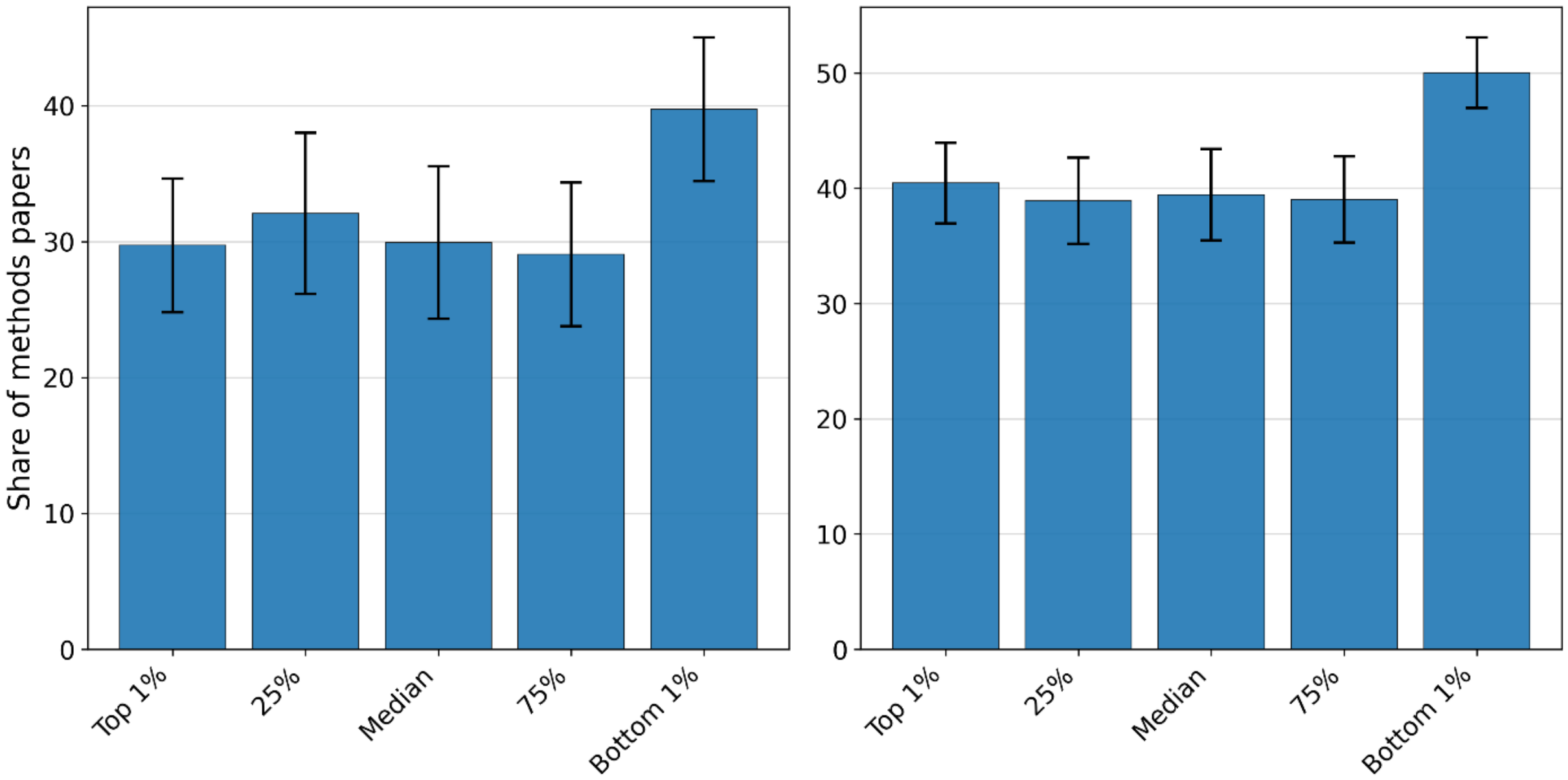

**Figure 1.** Share of methods papers across five stratified citation levels, ranging from the top 1% (highest citation impact, normalised by discipline and year) to the bottom 1% (lowest). Data is shown for Web of Science (left) and Scopus (right) for all years between 1980 and 2019.

**The great acceleration of the early 1990s.** Our analysis reveals a stark structural transition in how scientific knowledge is produced over time. Between 1980 and 2019, the share of methods papers doubled, rising from about 20% to 40% of the scientific literature over this period (Fig. 2). This growth does not follow a linear path but rather a classic S-shaped pattern, indicating a systemic transition rather than a gradual evolution. To pinpoint the precise timing of this shift, we applied segmented regression analysis to identify structural breaks—‘joinpoints’—in the trend (Fig. 3). The analysis reveals a pivotal discontinuity in the early 1990s. In the top 1% of papers for example, the prevalence of methods grew modestly (0.11 percentage points per year) prior to this period. But between 1988 and 1991, the growth rate surged more than thirty-fold—to 3.42 percentage points per year. Following this inflection point, the surge sustained an upward trajectory, implying that the shift toward methods is an ongoing feature of the research landscape rather than a transient spike. While a modest acceleration is also observed in the mid-2000’s within Scopus, it is not necessarily consistent with the WoS dataset and is not replicated in the segmented regression analysis—likely because Scopus does not extend back as far in time as WoS. Still, the early-1990s transition stands out clearly and robustly as the moment when methods became an increasingly central pillar of scientific research.

**Variations across disciplines**. The reorientation toward methods-driven science is a universal feature of science, though the precise timing in which disciplines experienced this shift varies (see figures in the Supplementary material and data section D). While the upward trajectory is omnipresent and the sigmoid temporal trajectory is common across disciplines, the timing and intensity of the acceleration slightly differs by discipline. Natural sciences display early onset—already accelerating in the late 1980s—whereas the medical and health sciences display a delayed onset, within the late 1990s and with a less profound infliction point.

Agricultural sciences follow a similarly strong upward trend in the 1990s, while engineering and technology illustrated earlier growth, in the late 1980s. Despite these differences, the overall pattern is remarkably consistent: overall agreement between WoS and Scopus databases confirms that science as whole—across disciplines, citation levels and over decades—has transitioned towards methods papers. The modest acceleration observed in the mid-2000's within Scopus does not alter the larger shift in methods dominance across science.

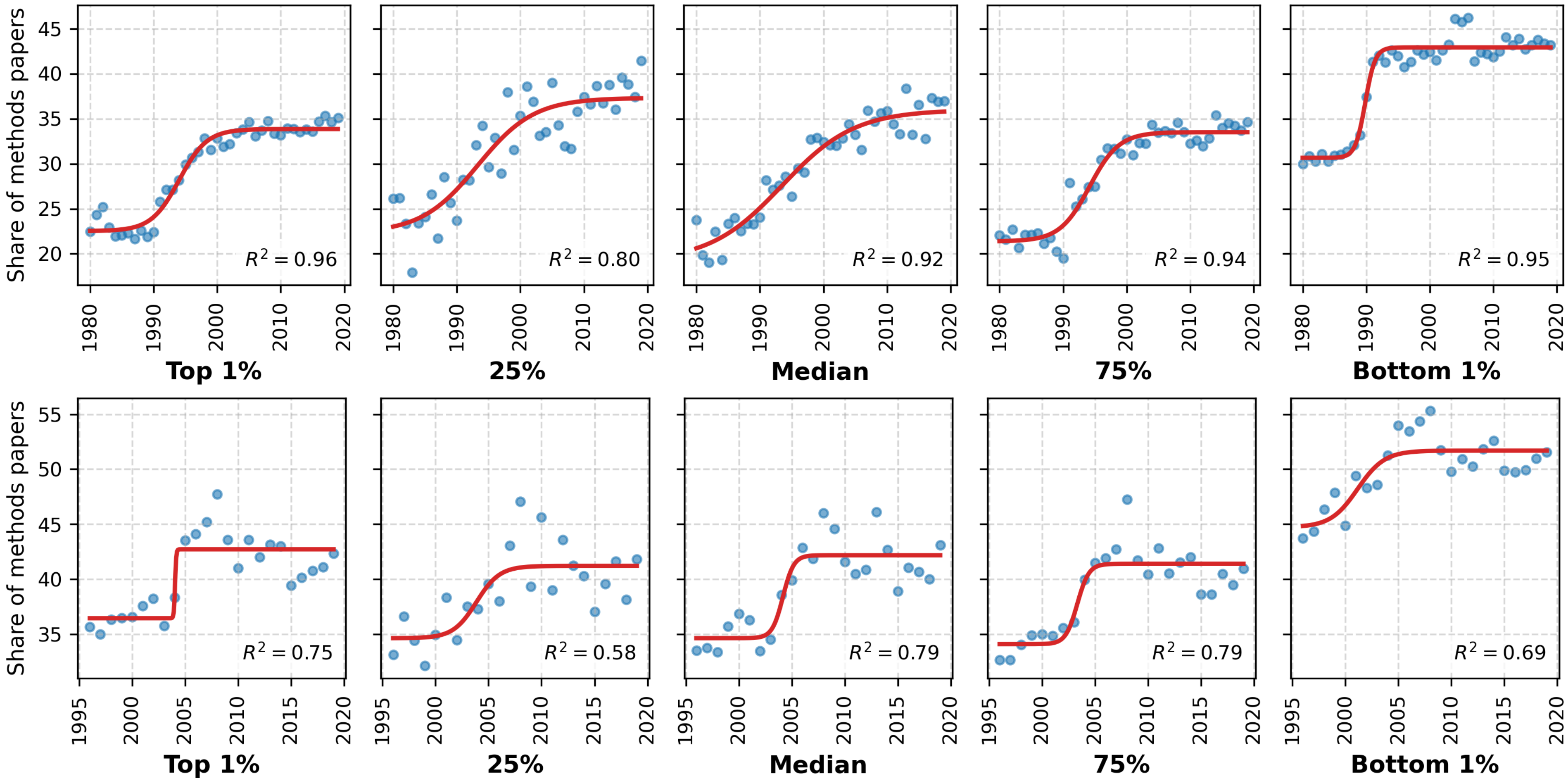

**Figure 2. Prevalence of methods papers by citation level across all disciplines, over time. Data is shown for Web of Science (top row) and Scopus (bottom row).**

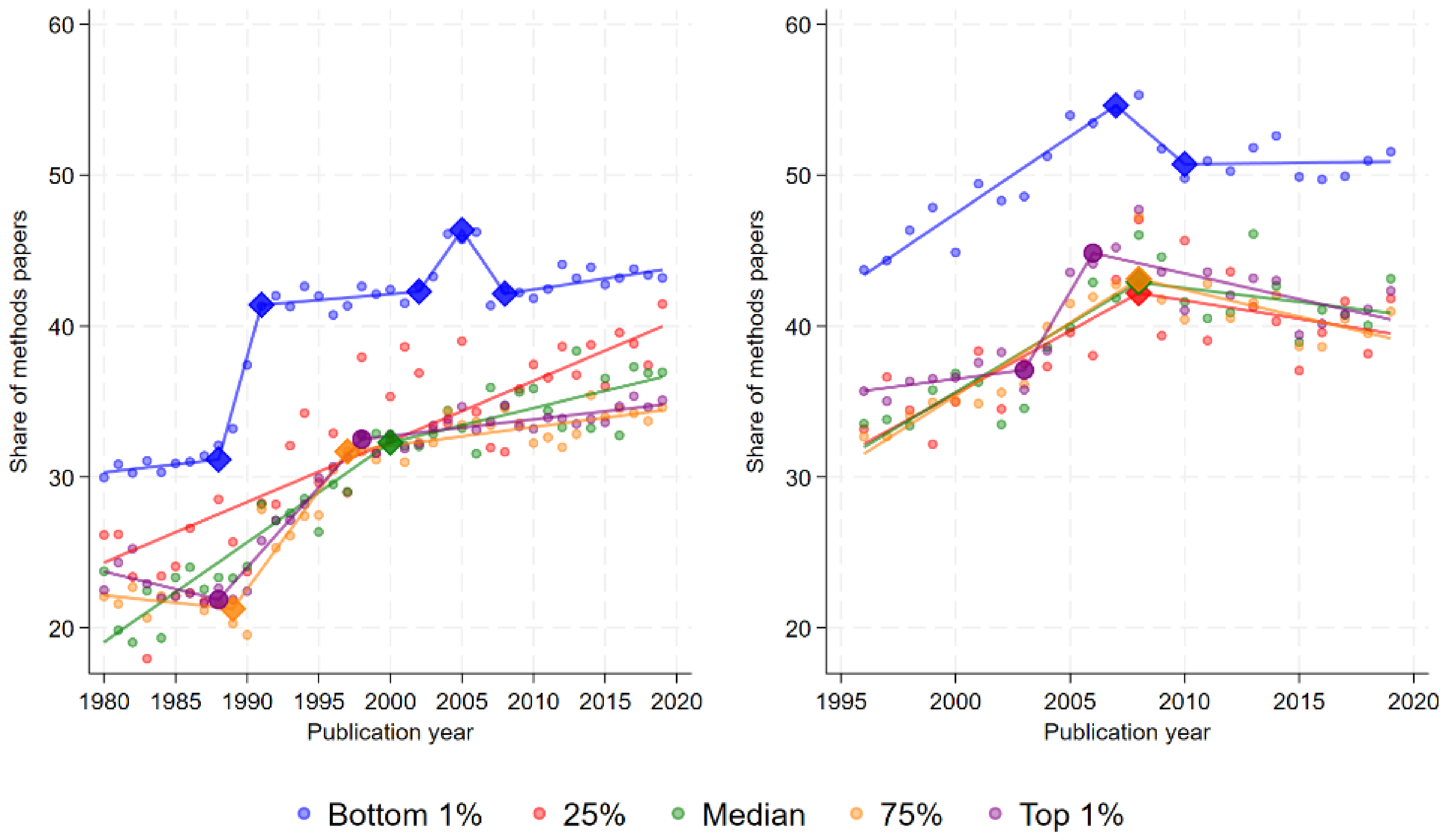

**Figure 3. Joinpoints in methods papers' prevalence across citation impact levels (for all disciplines), over time. Data is shown for Web of Science (left) and Scopus (right). Joinpoints are marked in the citation level lines.**

## 3 Discussion

Four key findings emerge from the analysis. First, we observe a sharp rise in the prevalence of methods papers over four decades, from the 1980s through the 2010s. Second, the early 1990s mark the beginning of the sharpest and most robust surge in methods papers. Third, this rise is consistent across citation impact levels, illustrating the shift is not confined to the scientific elite. Finally, while the observed upward trajectory is universal across all disciplines and data sources, its timing and intensity vary by discipline.

*Why have methods papers risen from 20% to 40%?* In explaining this methodological surge, we identify three interrelated structural forces:

**The computational and digital revolution**. Since the early 1990s, science has become fundamentally more computational, propelled by the rapid diffusion of affordable personal computers, early internet infrastructure and software ecosystems (7). These technologies enabled digital imaging, advanced statistical techniques, machine learning methods, algorithmic computations, high-throughput experimentation and unprecedented integration of instruments with computational tools—i.e. instrument-software integration at scale. Computer science helped lead the early transition by producing the general-purpose tools later adopted across disciplines. The PCR method (3) and the Human Genome Project, launched 1990, served as a catalyst, accelerating new automated instruments, sequencing platforms and statistical genetics—driving methods that diffused across biology and medicine (8). The combination of affordable computing power and sophisticated software transformed experimental design, data collection and analysis across the sciences (9).

**The rise of data-intensive science**. The shift toward data-driven science—big data—increased the demand for general-purpose methods and tools. From genomics to astronomy, big data expanded the demand for tools that could acquire, process, clean, analyse, store and visualise massive amounts of information and data (10, 11). The proliferation of statistical programmes and computational libraries (such as Python, R and MATLAB) further accelerated methodological innovation and lowered the barrier to implementing complex methods—with programmes like R free for researchers.

**Expanding interdisciplinarity**. Research has increasingly combined and recombined methodological approaches and techniques from different disciplines, favouring portable methods that can be re-used widely rather than methods tied to a single discipline (12). Because reusable methods serve larger communities across disciplinary boundaries, they tend to achieve higher levels of citation and adoption.

*Why the 1990s mark the turning point?* The structural break we identify in the 1990s coincides with several reinforcing technological and institutional transitions:

- Democratization of hardware: Personal computers from IBM and Apple became widely affordable for research labs by the early 1990s (13).
- Network effects: The World Wide Web was introduced in 1991, enabling unprecedented sharing of data, code and methodological protocols (14).
- Statistical software proliferation: Software packages like SPSS, Stata and R became increasingly accessible to non-specialists, opening advanced statistical methods to larger communities (15).
- Digital instrumentation: The digitalisation of lab instruments spurred digital microscopy, digital sequencing, digital detectors, digital imaging and other digital computer interfaces and automated data acquisition systems (9).
- Institutional catalysts: Mega-projects like the Human Genome Project (1990-2003) served as global accelerators, by demanding high-throughput sequencing and bioinformatics tools. These later diffused across biology and medicine (8).

The powerful combination of computing, statistics and digital instrumentation has made science exponentially more efficient, accelerating methodological innovation especially in the early 1990s. Across disciplines—from biology and medicine to physics—advances in computing have directly enabled new instruments, analysis methods and experimental designs, increasing scientific productivity (10).

*Why does the rise differ across disciplines?* While the results show consistent upward trends in methods papers across disciplines, the timing varies because of differences in technological adoption. In physics and chemistry, early growth in methods papers aligns with the spread of tools such as high-performance computing, simulation-based inference, instrument automation and advanced spectroscopy and imaging tools, which have enabled greater experimental advances. In biomedicine, the steepest surge corresponds to the rapid diffusion of high-throughput technologies—including PCR, automated DNA sequencing, fluorescence and digital microscopy, other imaging technologies and later microarrays. These reshaped experimental workflows and created sustained demand for reusable methods. In engineering and computer science, the rise began in the late 1980s, reflecting expanding algorithmic tools, numerical optimisation methods and computational architectures that laid the foundation for modern machine learning and software-driven tool development. These helped spur growth in other disciplines. In agriculture and environmental sciences, we observe a slower rise, primarily in the late 1990s and 2000s, consistent with delayed adoption of digital sensors, Geographic Information System (GIS) technologies, remote-sensing platforms and bioinformatics pipelines. Integrating satellite imagery, precision-agriculture tools and environmental-modelling software helped drive the methodological transformation in these disciplines (9).

Collectively, these dynamics can be conceptualized through the lens of the technological adoption lifecycle framework (16). The scientific community behaves as a social system that gradually adopts a new 'technology'—in this case, the convergence of computational methods, data-intensive science and interdisciplinary integration. As a result, multiple disciplines exhibit the classic S-curve of technological uptake: a phase of rapid acceleration in the 1990s (a structural break), followed by a maturation phase. This pattern signals a foundational shift where methods have evolved from specialized techniques into the generalized infrastructure of contemporary scientific practice—reflecting a new standard underlying how modern science is produced.

# 4 Conclusions

Over the past four decades, the scientific landscape has fundamentally shifted. Our study reveals that methods papers have moved from the periphery to the core of the scientific enterprise, becoming increasingly dominant not only in high-impact science but across all of science. The share of methods papers has doubled—rising from about 20% in 1980 to 40% in 2019—a trend that is robust across disciplines and citation impact levels. This surge accelerated sharply during the 1990s, aligning with the proliferation of digital technologies, the rise of data-intensive research and expanding interdisciplinary science. As a result, our findings indicate that today's influential papers are often not the advances themselves, but the reusable methods, tools and techniques that make those advances possible (6).

These findings point to the need to rethink how we conceptualize and carry out science. We need to recalibrate evaluation systems in science so that method architects are better credited for creating shared research infrastructure, better incentivised to publish in leading journals, better evaluated in tenure-track processes and better rewarded for innovations that enable downstream scientific advances. Beyond the immediate implications for funding, incentives and tenure, these results challenge the traditional model of science as a succession of theoretical paradigms and experimental findings. Instead, they highlight a system where progress is increasingly infrastructure-driven, and where the pace of progress is set by the availability of novel methods. In other words, “to make exciting advances, researchers rely on relatively unsung papers to describe experimental methods, databases and software” (1). This ‘methodisation’ of science we uncover implies that the boundary between ‘core research’ and ‘technical support’ is dissolving. Science policy must thus evolve to treat methods, software, databases and protocols not as secondary byproducts, but as primary scientific outputs that require sustained investment, distinct evaluation metrics and professional pathways for the architects of scientific infrastructure. Ensuring that the impact of these architects is rewarded is essential for sustaining the ecosystem of science.

While our analysis captures the digital transition up to 2019, it stops on the threshold of the AI era—an important frontier for future assessment. The widespread adoption of Large Language Models (LLMs) and AI agents introduces new dynamics. On one hand, their capacity to auto-generate code and design experimental protocols may lower barriers to creating new methods. On the other hand, AI can commoditize many scientific processes, potentially boosting scientific output overall. A critical question for the coming decade is which force will dominate: will AI catalyse a ‘hyper-methodisation’ of science, or simply amplify the volume of literature more broadly?

Our study has some limitations. While our classification model achieved high accuracy (>93%) and outperformed state-of-the-art baselines, text-based approaches carry inherent risks of misclassification (e.g. for a theory-focused paper introducing a statistical model). We mitigated this through a paired-journal training procedure, extensive robustness checks, inspection of the trained model’s operability, and analysing two databases separately. Also, our analysis relies on citation percentiles that cannot fully capture the scientific and societal influence of individual papers. Finally, we used data up to 2019 to exclude exceptional publication patterns associated with the Covid-19 pandemic and to ensure a citation window of at least five years that is common in bibliometrics to allow citations to accrue (17). If the 1990s marked the digitization of methods, the coming decade may signal their generative acceleration. AI-driven code generation, automated protocol design and synthetic data pipelines promise to reduce the cost of methodological innovation even further. Future research should examine whether these developments lead to a new ‘hyper-methodisation’ of science, where AI agents increasingly act as architects of scientific infrastructure. Ultimately, understanding this transformation is essential for navigating a new era of research—one in which the tools we build (and the intelligent systems that can help us build them) increasingly shape the questions we can ask.

## Acknowledgements

*Competing Interests*: There are no competing interests. *Ethical approval*: The study involves data compiled from the Web of Science (Clarivate) and Scopus (Elsevier), with no human experiments conducted and no ethical approval required. *Data availability*: Web of Science and Scopus publication-level data are subscription-restricted and subject to licensing agreements that prohibit redistribution. Aggregated discipline-level and yearly data are available by request.

## Methods and data

**Data collection and stratification**. We constructed a stratified corpus of journal articles indexed in the WoS between 1980 and 2019. We stratified the data annually into five distinct citation impact levels (18): the top 1%, 25% (24.5-25.5%), median (49.5–50.5%), 75% (74.5-75.5%) and bottom 1%. We restricted our analysis to the natural sciences, medical and health sciences, engineering and technology and agricultural sciences, domains with established empirical traditions, resulting in a dataset totalling 3,117,335 papers. To test the robustness of our findings, we collected a parallel validation set of papers from the Scopus database. WoS was selected as the primary dataset due to its historical depth, which is necessary to identify possible structural shifts occurring in the 1980s. Scopus provides comprehensive coverage for validation from the mid-1990s onward. Papers with missing titles, abstracts or publication years (<3%) were excluded.

**Machine learning classification framework**. We identified methods papers using a supervised machine-learning framework. To construct a balanced ground-truth dataset, we manually curated ten pairs of journals (20 in total), matching one ‘methods-oriented’ journal (e.g. *Nature Methods* or *Journal of Neuroscience Methods*) with one ‘general’ journal matched for scope and reputation (see Supplementary material and data section A for full list). This paired design forces the classifier to learn the lexical signals of method papers rather than using discipline or citation impact. We extracted titles and abstracts for all 180,454 papers in these journals over the covered time period. We vectorized the text using standard Term Frequency-Inverse Document Frequency (TF-IDF) and trained a Logistic Regression (LR) classifier (19). We benchmarked this TF-IDF+LR model against state-of-the-art alternatives, including SciBERT (20) and zero-shot/few-shot LLMs (GPT-4). Our model achieved the highest performance (accuracy: 93.9%; F1: 0.899) while maintaining high interpretability and reproducibility (see Supplementary material and data section B for comparative evaluation). We applied the trained model to the full WoS and Scopus datasets. The classifier assigned a probability score to each publication; papers exceeding a 0.5 confidence threshold were classified as ‘methods papers.’

**Statistical analysis**. Temporal trends were characterized using two complementary approaches. First, we fitted Sigmoid (logistic) functions to the time series to model the overall adoption trajectory. Second, to identify specific structural discontinuities, we applied segmented regression (joinpoint) analysis. Segmented regression is a statistical technique used to detect changes in trends within a time series. It represents the data as a sequence of linear segments, each with its own slope, allowing the model to capture periods of acceleration, deceleration or stability. The points at which the slope changes—called joinpoints—are not chosen in advance but are estimated directly from the data. Statistical tests evaluate whether adding a joinpoint significantly improves model fit, ensuring that detected changes are not random fluctuations. While the Sigmoid model may capture the overall shape of the transition, the segmented regression analysis allows us to statistically identify specific ‘break years’ where the acceleration rate significantly changes (21). The combination of both methods allows us to describe the *general nature* of the trend (adoption) and the *specific timing* of its acceleration.

# Supplementary material and data

## A. List of journal pairs

| Discipline | Type | Journal title | Publisher |
|---|---|---|---|
| Astronomy & astrophysics | Non-methods | *Astrophysical Journal* | American Astronomical Society |
| | Methods | *Astrophysical Journal, Supplement Series* | American Astronomical Society |
| Chemistry | Non-methods | *Physica Status Solidi - Rapid Research Letters* | John Wiley & Sons |
| | Methods | *Chemistry-Methods* | John Wiley & Sons |
| Earth sciences | Non-methods | *Chemical Geology* | Elsevier |
| | Methods | *Geostandards and Geoanalytical Research* | John Wiley & Sons |
| Ecology | Non-methods | *Ecology Letters* | John Wiley & Sons |
| | Methods | *Ecological Informatics* | Elsevier |
| Environmental science | Non-methods | *Global Change Biology* | John Wiley & Sons |
| | Methods | *Methods in Ecology and Evolution* | British Ecological Society |
| Materials science | Non-methods | *Journal of Colloid and Interface Science* | Elsevier |
| | Methods | *Surface and Interface Analysis* | John Wiley & Sons |
| Microbiology | Non-methods | *Emerging Microbes and Infections* | Taylor & Francis |
| | Methods | *Journal of Microbiological Methods* | Elsevier |
| Molecular biology | Non-methods | *Journal of Molecular Biology* | Elsevier |
| | Methods | *Methods* | Elsevier |
| Neuroscience | Non-methods | *Journal of Neuroscience* | Society for Neuroscience |
| | Methods | *Journal of Neuroscience Methods* | Elsevier |
| Physics | Non-methods | *Physical Review Letters* | American Physical Society |
| | Methods | *Review of Scientific Instruments* | American Institute of Physics |

## B. Classification models' evaluation

**TF–IDF + Logistic regression**
Delivers strong performance with high interpretability, minimal computational cost and enhanced reproducibility.

- Accuracy: 0.939
- Recall: 0.873
- F1: 0.899

**SciBERT + Logistic regression**
Comparable performance to TF–IDF+LR but considerably more computationally expensive and less interpretable.

- **Accuracy:** 0.935
- **Recall:** 0.875
- **F1:** 0.893

**LLM (GPT-5) zero-shot**
Non-deterministic; averaged over 10 runs (majority voting). Results were close to chance.

- **Accuracy:** 0.576
- **Recall:** 0.541
- **F1:** 0.558

**LLM (GPT-5) few-shot**
Non-deterministic; averaged over 10 runs (majority voting). Ten paired 'hallmark' papers, half manually classified as 'methods' and half classified as 'non-methods', were included into the model (listed below). Slight improvement over the zero-shot baseline but still outperformed by the TF-IDF+LR model.

- **Accuracy:** 0.637

- **Recall:** 0.629
- **F1:** 0.633

## C. Papers used by the few-shot learning model

* Methods paper: *(Tool for molecular construction – organocatalysis; Benjamin List, David MacMillan; 2000; Chemistry)*
Ahrendt, K. A., Borths, C. J., & MacMillan, D. W. C. New strategies for organic catalysis (2000). The first highly enantioselective organocatalytic Diels-Alder reaction. *Journal of the American Chemical Society, 122* (17), 4243-4244.

Non-methods paper: *(Mapping of ribosomes; Venkatraman Ramakrishnan, Thomas A. Steitz, Ada E. Yonath; 2000; Chemistry)*
Wimberly, B., Brodersen, D., Clemons, W. ... V. Ramakrishnan (2000). Structure of the 30S ribosomal subunit. *Nature, 407*, 327–339.

* Methods paper: *(Stimulated emission depletion microscopy; Stefan W. Hell; 1994; Chemistry)*
Hell, S. W. & Wichmann, J. (1994). Breaking the diffraction resolution limit by stimulated emission: stimulated-emission-depletion fluorescence microscopy, *Optics Letters, 19*, 780-782.

Non-methods paper: *(Structure of ATP synthase; John E. Walker; 1994; Chemistry)*
Abrahams, J., Leslie, A., Lutter, R., & Walker, J. (1994). Structure at 2.8 Â resolution of $F_1$-ATPase from bovine heart mitochondria. *Nature, 370*, 621–628.

* Methods paper: *(Click chemistry; K. Barry Sharpless, Morten Meldal; 2001; Chemistry)*
Kolb, H. C., Finn, M. G., & Sharpless, K. B. (2001), Click chemistry: Diverse chemical function from a few good reactions. *Angewandte Chemie – International Edition, 40*, 2004-2021.

Non-methods paper: *(Molecular basis of eukaryotic transcription; Roger D. Kornberg; 2001; Chemistry)*
Cramer, P., David A. Bushnell, D. A., & Kornberg, R. D. (2001). Structural basis of transcription: RNA Polymerase II at 2.8 Ångstrom Resolution. *Science, 292*, 1863-1876.

* Methods paper: *(Velocity selective coherent population trapping method; Claude Cohen-Tannoudji; 1988; Physics)*
Aspect, A., Arimondo, E., Kaiser, R., Vansteenkiste, N., & Cohen-Tannoudji, C. (1988). Laser cooling below the one-photon recoil energy by velocity-selective coherent population trapping. *Physical Review Letters, 61*(826).

Non-methods paper: *(Giant magnetoresistance; Albert Fert, Peter Grünberg; 1988; Physics)*
Baibich, M. N., Broto, J. M., Fert, A. et al. (1988). Giant magnetoresistance of (001)Fe/(001)Cr magnetic superlattices. *Physical Review Letters*, 61(2472), 2472-2475.

* Methods paper: *(Polymerase chain reaction (PCR) method; Kary B. Mullis; 1985; Chemistry)*
Saiki, R. K., Scharf, S., Faloona, F., Mullis, K. B., Horn, G. T., Erlich, H. A., & Arnheim, N. (1985). Enzymatic amplification of β-Globin genomic sequences and restriction site analysis for diagnosis of sickle cell anemia, *Science, 230*(4732), 1350-1354.

Non-methods paper: *(Fullerenes; Robert F. Curl Jr., Sir Harold W. Kroto, Richard E. Smalley; 1985; Chemistry)*
Kroto, H., Heath, J., O'Brien, S. et al. (1985). C60: Buckminsterfullerene. *Nature, 318*, 162–163.

* Methods paper: *(Vitrification water technique for electron cryo-microscopy; Jacques Dubochet; 1984; Chemistry)*
Adrian, M., Dubochet, J., Lepault, J., & McDowall, A. W. (1984). Cryo-electron microscopy of viruses, *Nature, 308*, 32-36.

Non-methods paper: *(Quasicrystals; Dan Shechtman; 1984; Chemistry)*
Shechtman, D., Blech, I., Gratias, D., & Cahn, J.W. (1984). Metallic phase with long range orientational order and no translation symmetry. *Physical Review Letters, 53*(20), 1951-1954.

* Methods paper: *(Scanning tunneling microscope; Gerd Binnig, Heinrich Rohrer; 1982; Physics)*
Binnig, G., Rohrer, H., Gerber, Ch., & Weibel, E.. (1982). Surface studies by scanning tunneling microscopy. *Physical Review Letters*, *49*, 57.

Non-methods paper: *(Fractional quantum Hall effect; Horst L. Störmer, Daniel C. Tsui, Robert B. Laughlin; 1982; Physics)*
Tsui, D. C., Stormer, H. L., & Gossard A. C. (1982). Two-dimensional magnetotransport in the extreme quantum limit. *Physical Review Letters, 48*, 1559-1562.

* Methods paper: *(Site-directed mutagenesis; Michael Smith; 1982; Chemistry)*
Zoller, M. J., & Smith, M. (1982). Oligonucleotide-directed mutagenesis using M13-derived vectors: an efficient and general procedure for the production of point mutations in any fragment of DNA. *Nucleic Acids Research, 10*(20), 6487–6500.

Non-methods paper: *(Structure of a photosynthetic reaction centre; Johann Deisenhofer, Robert Huber, Hartmut Michel; 1982; Chemistry)*
Michel, H. (1982). Three-dimensional crystals of a membrane protein complex. *Journal of Molecular Biology, 158*(3), 567–572.

* Methods paper: (*Method for identifying function of single ion channels in cells; Erwin Neher, Bert Sakmann; 1981; Physiology/medicine)*
Hamill, O.P., Marty, A., Neher, E. et al. (1981). Improved patch-clamp techniques for high-resolution current recording from cells and cell-free membrane patches. *Pflügers Archiv, 391*, 85–100.

Non-methods paper: *(Cultivation of embryonic stem cells; Sir Martin J. Evans; 1981; Physiology/medicine)*
Evans, M., & Kaufman, M. (1981). Establishment in culture of pluripotential cells from mouse embryos. *Nature, 292*, 154–156.

* Methods paper: *(Chirally catalysed oxidation reactions; K. Barry Sharpless; 1980; Chemistry)*
Katsuki; T., & Sharpless, K. B. (1980). The first practical method for asymmetric epoxidation. *Journal of the American Chemical Society*, *102*(18), 5974–5976.

Non-methods paper: *(Ubiquitin-mediated protein degradation; Aaron Ciechanover, Avram Hershko, Irwin Rose; 1980; Chemistry)*
Hershko, A., Ciechanover, A., Heller, H., Haas, A. L., & Rose, I. A. (1980). Proposed role of ATP in protein breakdown: conjugation of protein with multiple chains of the polypeptide of ATP-dependent proteolysis. *Proceedings of the National Academy of Sciences of the United States of America*, *77*(4), 1783-1786.

# D. Disciplinary figures

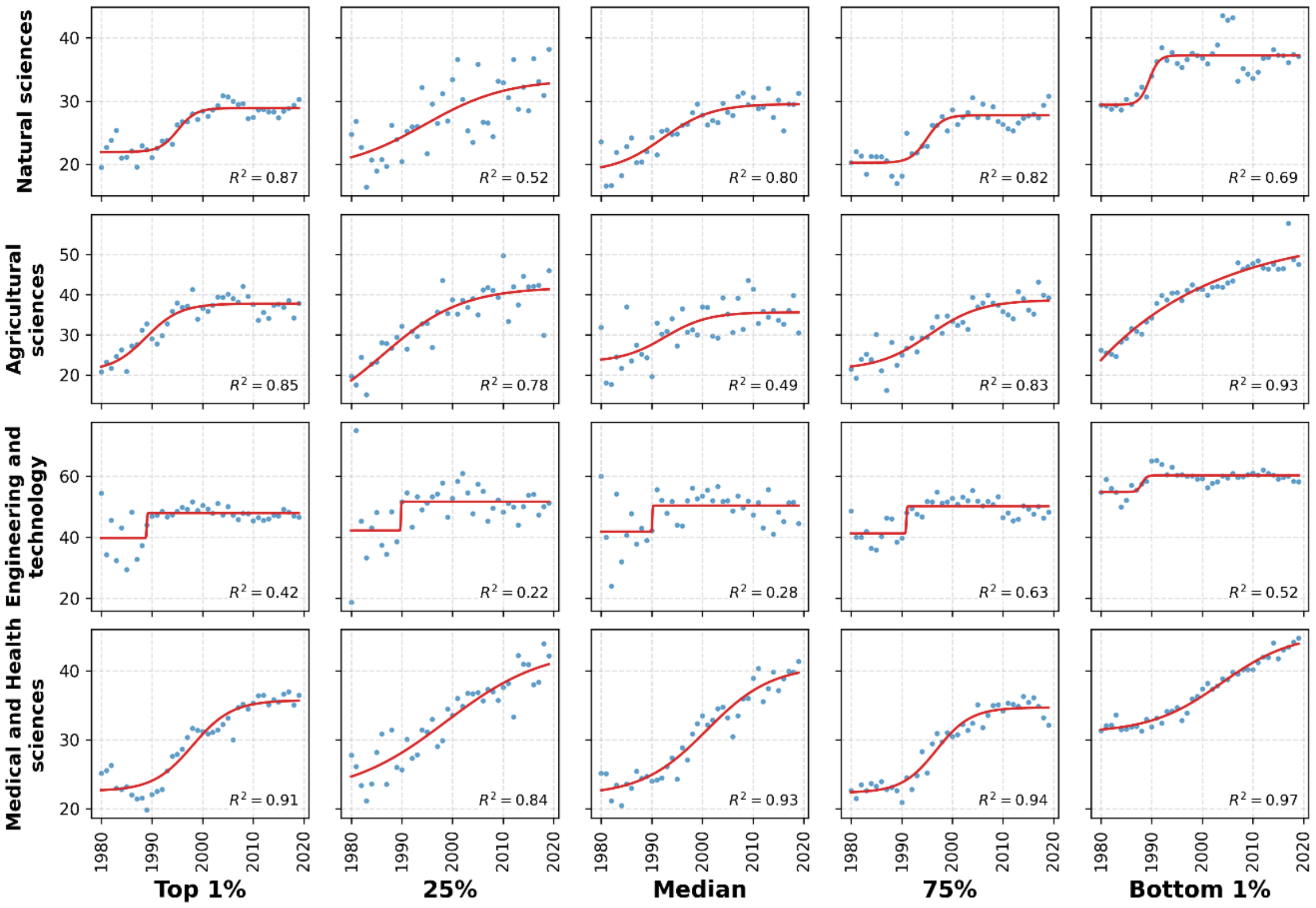

**Appendix Figure 1. Share of methods papers across citation impact levels and disciplines, over time (WoS data).**

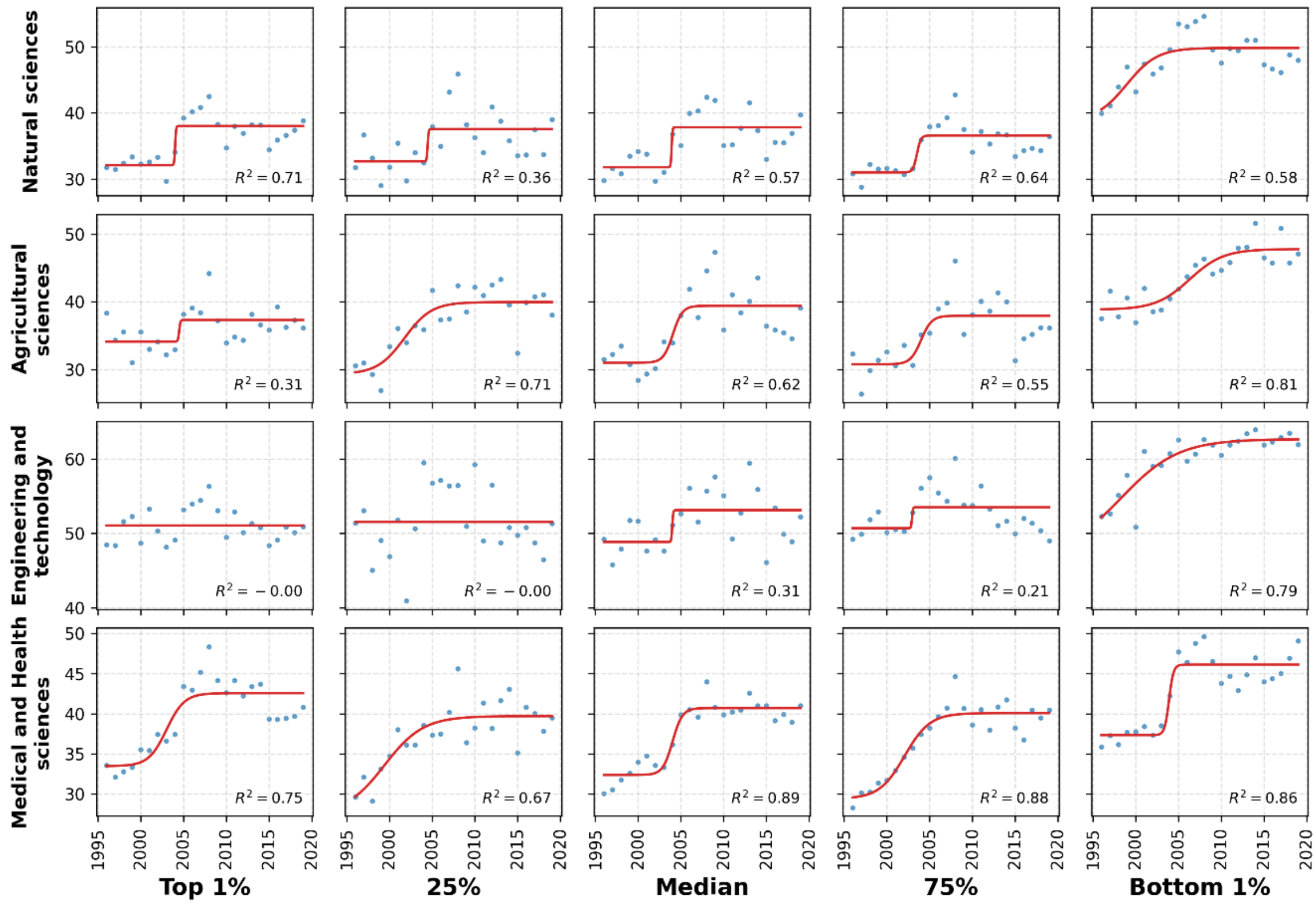

**Appendix Figure 2. Share of methods papers across citation impact levels and disciplines, over time (Scopus data).**

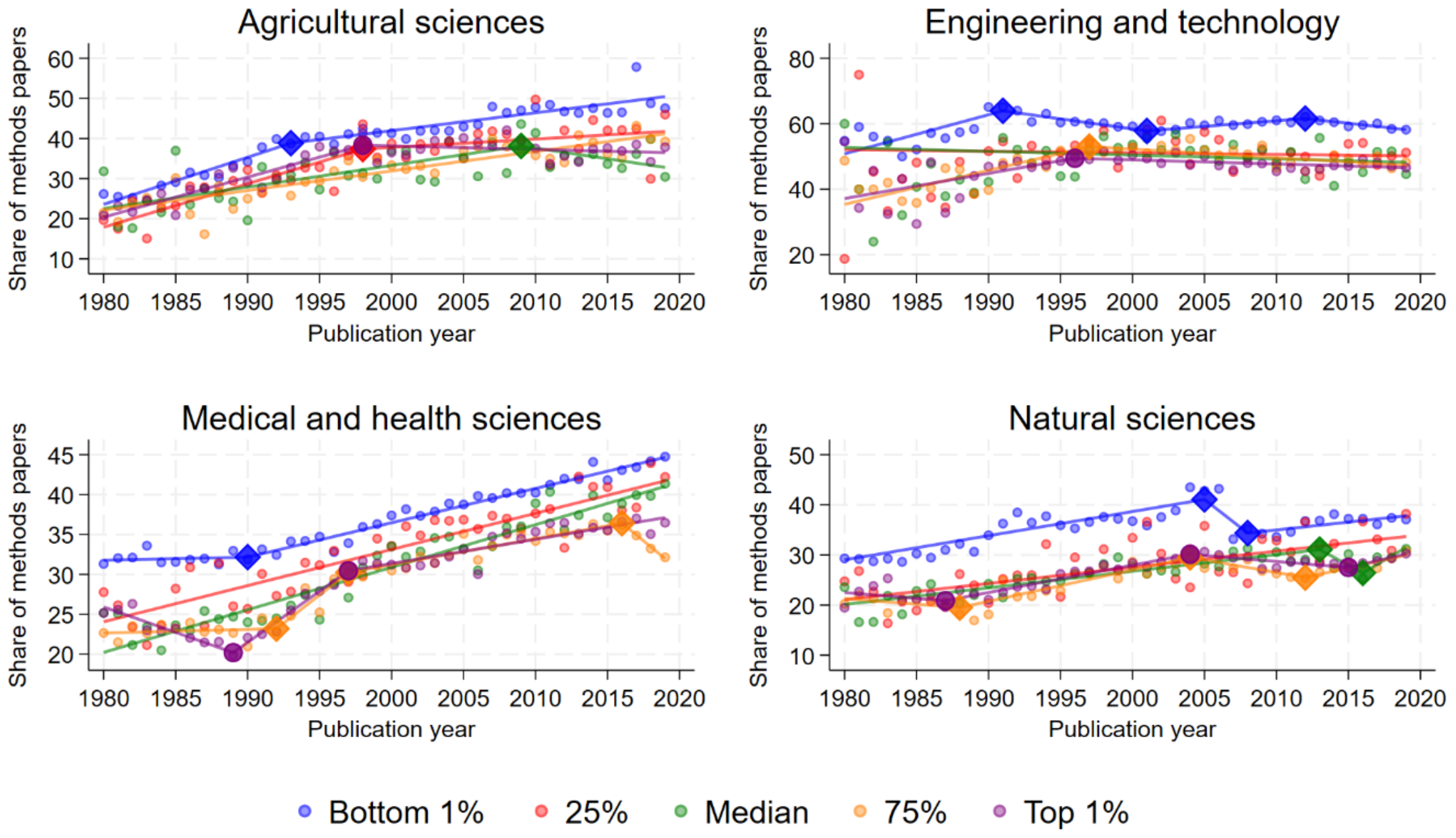

**Appendix Figure 3. Joinpoints in methods papers' prevalence across citation impact tiers and disciplines, over time (WoS data). Joinpoints are marked in the citation level lines.**

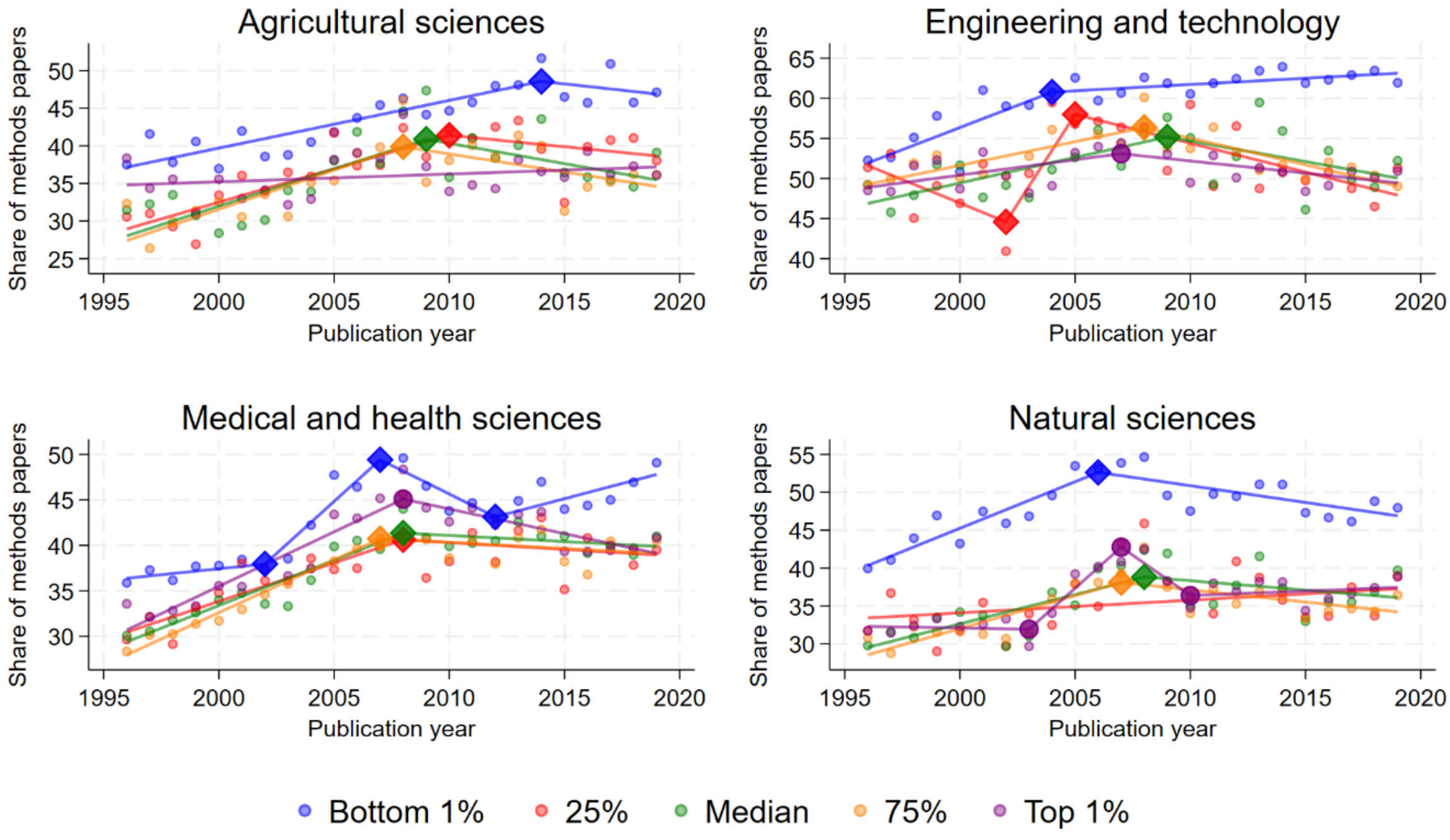

**Appendix Figure 4. Joinpoints in methods papers' prevalence across citation impact tiers and disciplines, over time (Scopus data). Joinpoints are marked in the citation level lines.**